# Developing Critical Thinking in Second Language Learners: Exploring Generative AI like ChatGPT as a tool for Argumentative Essay Writing

Simon Suh, Jihyuk Bang, Ji Woo Han

[Abstract]

This study employs the Paul-Elder Critical Thinking Model and Tan's argumentative writing framework to create a structured methodology. This methodology, ChatGPT Guideline for Critical Argumentative Writing (CGCAW) framework, integrates the models with ChatGPT's capabilities to guide L2 learners in utilizing ChatGPT to enhance their critical thinking skills. A quantitative experiment was conducted with 10 participants from a state university, divided into experimental and control groups. The experimental group utilized the ChatGPT Guideline for Critical Argumentative Writing (CGCAW) framework, while the control group used ChatGPT without specific guidelines. Participants were tasked with writing a short argumentative essay within a 40-minute timeframe. The essays were evaluated by three assessors: ChatGPT, Grammarly, and a course instructor from the state university, each using different grading criteria aligned with the CGCAW framework. Results indicated that the experimental group showed improvements in clarity, logical coherence, and use of evidence, demonstrating ChatGPT's potential to enhance specific aspects of argumentative writing. However, the control group performed better in overall language mechanics and articulation of main arguments, indicating areas where the CGCAW framework could be further refined. This study highlights the need for further research to optimize the use of AI tools like ChatGPT in L2 learning environments, to enhance critical thinking and writing skills.

# 1. [Introduction]

The process of language learning involves multifaceted skill sets such as reading, listening, speaking, and writing (Bimpong, 2020). Among these, writing stands out for its unique ability to promote not only linguistic proficiency but also critical thinking skills (Astuti et al., 2020; Wu, 2021). During the writing process, learners enhance their writing by analyzing and evaluating the given information through the application of critical thinking skills (Wu, 2021). A study conducted by Hasnawati et al. (2023) supports this finding, revealing that students who struggled with developing and writing their theses often lacked critical thinking skills. This suggests that without practicing critical thinking skills, students are unable to integrate evidence effectively and construct their arguments.

It is believed that critical thinking fosters the interplay between linguistic features and content construction, leading to successful writing ability (Wu, 2021; Raj et al., 2022). Argumentative writing requires the writer to analyze and evaluate content knowledge, develop a clear position, and present a coherent argument (Wu, 2021). This means that if learners are unable to activate their critical thinking skills, they will be unable to develop their ideas and writing (Tosuncuoglu, 2018). In other words, critical thinking is essential at all stages of writing in language learning. As Raj et al. (2022) argued, writing involves investigating, evaluating, interpreting, synthesizing information, and applying creative thought to solve problems. The process requires critical thinking because writing an argumentative essay is ultimately a direct reflection of the writer's thoughts (Yundayani, 2021). Thus, the relationship between critical thinking and crafting argumentative writing is symbiotic, promoting the natural progression of language learning for second language learners.

ChatGPT can function as a versatile and valuable dialogue partner for language learners (Kohnke, 2023), enhancing their critical thinking skills by challenging their way of thinking during the writing process (Van Rensburg, 2024). A study conducted by Guo et al. (2023) explored the use of ChatGPT in higher education and found that students reported it provided diverse perspectives and challenged their existing ways of thinking by broadening their understanding, enhancing analysis skills, and reassessing their preconceptions. ChatGPT achieved this by presenting alternative viewpoints on topics, prompting students to consider different angles and arguments they might not have previously encountered. For instance, when discussing a controversial issue, ChatGPT could generate and help students form arguments for and against the topic. Additionally, ChatGPT facilitated critical thinking by asking probing questions and suggesting further areas of research, which encouraged students to develop a deeper analysis. By offering counterarguments and highlighting potential biases in students' initial assumptions, ChatGPT prompted them to reassess their preconceptions and develop a more comprehensive understanding of the material. This interaction not only enhanced their analytical skills but also fostered a more open-minded and critical approach to their studies. In the multi-stage essay writing process, which includes interpretation, analysis, evaluation, inference, and explanation, ChatGPT assisted users in analyzing information and evaluating complex concepts. Interaction with ChatGPT helped students develop a more critical approach to information analysis. As language learners employed ChatGPT throughout their writing process, they were able to effectively practice higher-order thinking skills, thereby fostering the development of critical thinking skills.

Although direct research linking critical thinking and ChatGPT in language learning is limited, ChatGPT has shown capabilities that can enhance critical thinking. It can generate diverse perspectives, refine writing through iterative prompts, and validate information. These features challenge students' thoughts and stimulate the development of their critical thinking skills in writing. Building upon these insights, this research aims to (a) investigate the effectiveness of ChatGPT in developing critical thinking skills for second language learners through argumentative writing; (b) create a framework for second language learners to utilize ChatGPT in developing critical thinking skills; and (c) compare the quality of writing between control and experimental groups to evaluate the effectiveness of the framework.

## 2. [Literature Review]

### 2.1.1 Elements of Critical Thinking and Paul-Elder Critical Thinking Model

In the context of learning, Paul et al. (2006) identify eight elements of thinking: purpose, questions, information, inferences, concepts, implications, assumptions, and point of view. Bekele et al. (2021) further explain that these elements empower individuals to identify issues, recognize significant relationships, deduce conclusions from available information, evaluate evidence, and draw conclusions. This process fosters the development of an individual's critical thinking skills.

The process of developing critical thinking requires active engagement with several key elements. Initially, individuals must establish a clear purpose for their thinking, setting specific objectives to guide their analysis. They then pose relevant questions to further explore the topic and collect necessary information to address these questions (Nappi, 2017). As they process the information, individuals make inferences, draw logical conclusions, and use concepts such as mental frameworks to organize their understanding. They also consider the implications of their conclusions, evaluating potential consequences and applications of their reasoning. Throughout this process, individuals critically examine their assumptions and challenge their beliefs, which strengthens their perspectives. By recognizing their point of view, they cultivate openness to alternative viewpoints, which helps them develop a more robust perspective on the issue (Southworth, 2022).

The Paul-Elder Critical Thinking Model (PECTM) (2006) offers valuable insights into the critical thinking process by illustrating how individuals can enhance their quality of thinking. By taking charge of innate thinking standards and aligning them with the eight elements of thinking, individuals cultivate eight intellectual traits. This nurturing of intellectual traits enables the progressive development of critical thinking skills through continuous practice, as illustrated in Figure 1.

In our research, of many other researchers such as Al-Ghadouni (2021), Chukwuere (2024) and Raj et al. (2022) who suggested a critical thinking model, we selected the Paul-Elder Critical Thinking (PECTM) Model shown in Figure 1 below due to its international recognition and comprehensive approach to evaluating and enhancing critical thinking skills (CriticalThinking.org).

We will utilize the PECTM model to investigate and analyze how individuals develop critical thinking capabilities while writing argumentative essays. The elements of reasoning—purpose, question, information, inference, concepts, assumptions, implications, and point of view—will guide our analysis. We will assess how clearly students articulate their essay's purpose and frame their research questions. The quality and relevance of the information they gather, as well as the soundness of their inferences and conclusions, will be evaluated. Additionally, we will examine their understanding and application of key concepts, the recognition and justification of their assumptions, and their consideration of the implications of their arguments. Finally, we will review how well students present and integrate multiple points of view. By focusing on these elements, we aim to provide structured feedback that enhances students' critical thinking skills in their argumentative writing.

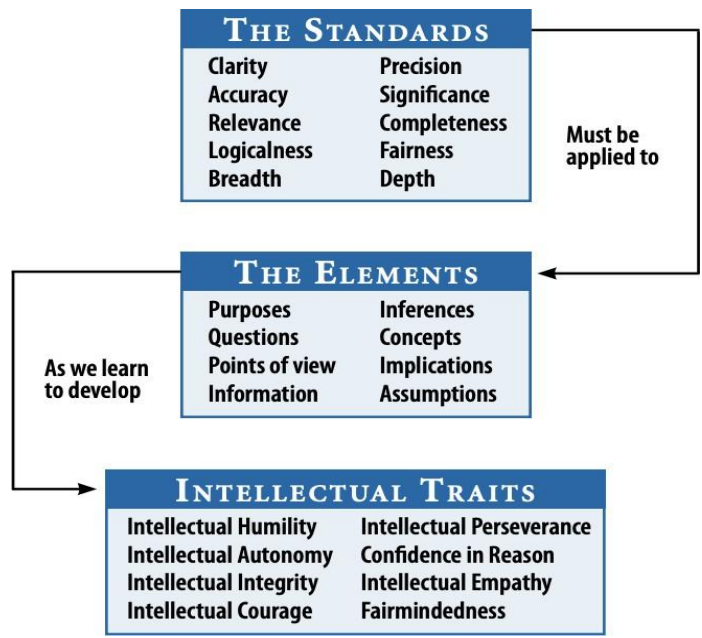

Figure 1: Paul-Elder Critical Thinking Model (Paul et al., 2006, p.21)

**2.1.2 Argumentative Writing: Importance to Second Language Learners**

Argumentative writing is used to argue for or against a specific prompt by providing efficient reasoning and evidence (Purdue University, 2018). Crafting an effective argumentative essay is a multi-step process. Writers must research a topic, critically assess evidence, and establish a clear, defendable position (Purdue University, 2018). Essentially, successful argumentative writing demands strong analysis, effective evaluation, and logical reasoning regarding the writer's position on the topic.

Clearly articulating a stance is crucial for impactful argumentative writing. This not only requires fluency in the language but also the ability to organize ideas persuasively, which is precisely why argumentative writing presents itself as a significant challenge for second language (L2) learners. Unlike other forms of writing, argumentative writing requires a deeper understanding of the language and a higher level of critical thinking to analyze problems, evaluate source biases, and construct logical arguments (Pei et al., 2017), which presents itself as a significant challenge for second language learners (L2 learners).

Elements of Paul-Elder Critical Thinking Model and PECTM model application in argumentative essays encourages L2 learners to explore a topic, collect relevant evidence, and formulate a well-organized argument (ZR, 2022). Most importantly, it provides a platform for practicing critical thinking skills such as analyzing information, evaluating sources, and developing logical arguments, thus aiding L2 learners in enhancing their critical thinking skills through the process of argumentative writing.

### 2.1.3 PECTM Model Application in Argumentative Essay

Tan (2023) argued that there are four major elements of argumentative writing. His investigation focused on how these elements integrate with the six elements of thinking outlined in the PECTM model. Within this framework, Tan identified two elements, purposes and questions, from the PECTM model that were difficult to apply to argumentative writing and chose to exclude them from his analysis. He discovered a symbiotic relationship between the elements of argumentative writing and those from the PECTM model, as detailed in Table 1 of the study.

| Major Parts of an Argumentative Essay | Elements of Critical Thinking |
|---|---|
| **Position(s):** central position and alternative positions, each with main claim and supporting claims | **Points of view** |
| **Explanations:** explanations for supporting claims | **Concepts** **Assumptions** |
| **Evidence:** personal experiences, statistics, anecdotes, research findings, etc. | **Information** |
| **Conclusion:** restatement of the position(s), inferences, and implications | **Points of view** **Inferences** **Implications** |

Table 1: Connections between Major parts of an L2 Argumentative Essay and Elements of Critical Thinking

(Tan, 2023, p.6)

The process of writing an argumentative essay follows a structured sequence, as outlined by Tan, who emphasizes that the thinking processes of individuals are shaped by specific purposes, questions, and assumptions. These factors guide them to adopt particular viewpoints. In essence, effective argumentative writing enables writers to articulate and defend their positions, reflecting their unique perspectives. This initial step is followed by a comprehensive explanation phase, during which writers engage with the underlying concepts and assumptions that inform their arguments. Additionally, the critical task of gathering and analyzing evidence helps solidify their claims, thus establishing a robust foundation for their arguments. In the essay's conclusion, writers restate their positions, draw logical inferences, and explain the broader implications of their arguments.

By employing the PECTM model as a guiding framework, Tan effectively demonstrated how writers can navigate the complexities of argumentative writing through a structured approach. This successful integration of the internationally recognized PECTM model led us to adopt Tan's approach. Our research aims to investigate how Tan's methodical strategy enhances writing proficiency and fosters critical thinking skills among L2 learners through the use of ChatGPT.

## 2.2 Importance of ChatGPT and its Feedback in Argumentative Writing

Constructing a high-quality argumentative essay requires consideration of the conversational, structural, and linguistic aspects (Su et al., 2023). To achieve this, students need to sift through large amounts of data and research papers to find relevant information, a highly time-consuming process. ChatGPT can assist by analyzing and summarizing vast amounts of information more effectively and efficiently (Esmaeil et al., 2023). ChatGPT demonstrates improved natural language comprehension, efficiency, and accuracy in responding to queries, as well as enhanced adaptability (Su et al., 2023). Given these capabilities, ChatGPT holds promising potential to assist students in overcoming the conversational, structural, and linguistic challenges of argumentative writing.

ChatGPT is considered capable of supporting students in writing by helping with the major elements of the argumentative essay: Position, Explanation, Evidence, and Conclusion (Tan, 2023). As learners substantiate their arguments, ChatGPT can provide feedback on each category, identifying potential errors, inconsistencies, and gaps in the writer's position, explanation, evidence, and conclusion. Research by Manhapatra (2024) supports this, finding that ChatGPT was able to provide formative feedback that leads to reflection, self-regulation, self-monitoring, and revision. By utilizing specific features of ChatGPT, such as its ability to analyze text, generate suggestions for improvement, and offer examples of effective writing, it supports writers by providing appropriate directions related to content and organization. This process acts as a catalyst for enhancing the academic writing skills of English Second Language (L2) learners. By emphasizing the significance of ChatGPT's feedback and its profound impact on enhancing L2 argumentative writing, it becomes evident that this innovative tool holds immense potential in academic writing. Specifically, ChatGPT facilitates these benefits through real-time feedback, targeted suggestions, and examples of best practices in writing. This feedback not only helps in refining the essay's structure but also encourages critical thinking, enabling students to critically analyze their arguments and evidence.

According to Mondal (2023), ChatGPT was strategically employed by L2 learners to assess its overall impact on argumentative writing. The study focused on specific components of argumentative writing, including hypothesis generation, literature review, reference generation, methodology selection, and data collection. By asking questions within these specific components, students were able to interact with ChatGPT in each category, analyzing, critiquing, and utilizing the generated text to enhance their writing. Mondal's research asserts that to effectively use ChatGPT for writing argumentative essays, specific keywords or research topics must be provided for efficient output. Mondal suggested the strategy for getting better responses from ChatGPT as this: Although ChatGPT can remember previous conversations, its memory has certain limitations. Therefore, authors are advised to divide the literature review topic into smaller segments. They can then generate text for each segment individually and combine them afterward to ensure coherence and completeness. Through this interaction, results showed that ChatGPT had a significant positive influence on the writer's argumentative writing.

## 2.2.1 ChatGPT's Impact in Argumentative Writing through Hattie & Timperley Feedback Model (2007)

Feedback is regarded as one of the most crucial processes in students' academic learning (Gary et al., 2022). Noroozi et al. (2016) claimed that peer feedback significantly enhances the quality of students' written argumentative essays. This idea is supported by the Hattie & Timperley model (2007), which identifies four levels of feedback addressing different aspects of student learning: task, process, self-regulation, and self (shown in Figure 2). This model explains that each level helps students engage efficiently in educational activities through feedback. By providing feedback for each specific criterion, students gain insight into the benchmarks against which they can evaluate their writing, enhancing their understanding of their progress toward successful writing in various genres, including argumentative writing (Hattie & Timperley, 2007).

The first step of the model involves clarifying goals, enabling learners to establish clear objectives by interacting with ChatGPT. Through this interaction, learners can ask ChatGPT to help define the scope of their writing, set specific goals for their argumentation, and receive guidance on structuring their work effectively. ChatGPT can provide tailored advice on how to outline their essay, develop a thesis statement, and identify key points to support their arguments. In the second step, providing effective feedback, learners can utilize ChatGPT to identify areas for improvement in their writing. By inputting their drafts into ChatGPT, they receive detailed feedback on various aspects such as grammar, coherence, logical flow, and the strength of their evidence. ChatGPT offers suggestions for enhancing the clarity and persuasiveness of their arguments. This includes pointing out inconsistencies, recommending more precise vocabulary, and suggesting better ways to organize their thoughts. The third step, the student response phase, allows active engagement with the feedback generated by ChatGPT. In this phase, learners interact with the feedback provided, which is categorized across four levels: task, process, self-regulation, and self.

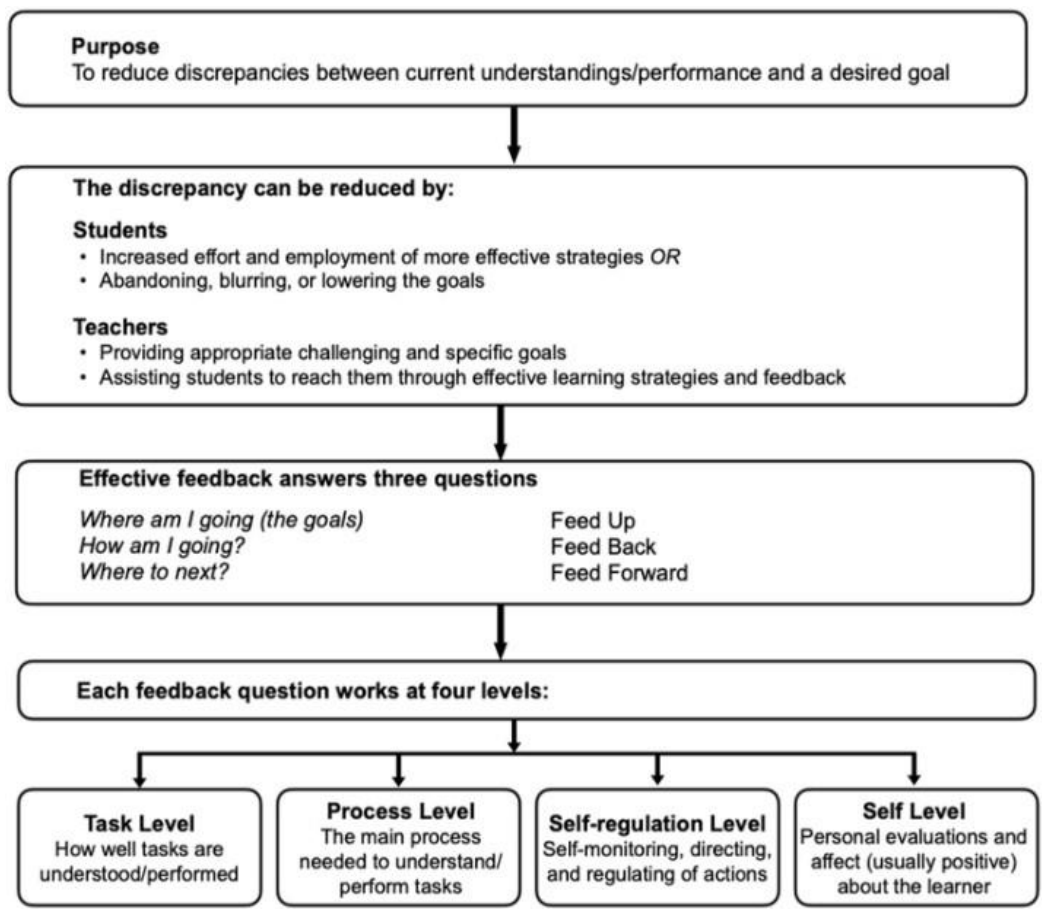

Figure 2: A Model of Feedback to Enhance Learning (Hattie & Timperley, 2007)

The feedback model proposed by Hattie and Timperley (HT) offers a valuable framework for structuring strategies aimed at utilizing ChatGPT to enhance critical thinking and improve writing skills, particularly for L2 learners. This is achieved by providing detailed, task-specific feedback that clarifies misunderstandings, guides learning, and encourages self-regulation. Our research focuses on fostering critical thinking skills among L2 learners through argumentative writing. Therefore, we have adopted the HT model as our guiding framework to explore the feedback process.

By integrating the HT model with Tan's framework of argumentative writing, we can demonstrate how ChatGPT's feedback contributes to developing critical thinking skills and improving writing abilities. The HT model's four levels—task, process, self-regulation, and self—are aligned with the elements of argumentative writing: Position, Explanation, Evidence, and Conclusion. At the task level, ChatGPT provides specific feedback on the content, helping learners understand their mistakes and improve their arguments. At the process level, it offers strategies for structuring and organizing their essays. For self-regulation, ChatGPT encourages students to monitor and reflect on their progress, fostering an independent and critical approach to their writing. Finally, at the self-level, ChatGPT helps build confidence by acknowledging improvements and providing positive reinforcement.

Through this approach, we aim to uncover the potential feedback that ChatGPT can provide to L2 students, thus enhancing their abilities in argumentative writing. The interaction with ChatGPT not only improves the technical aspects of writing but also promotes critical thinking by encouraging students to analyze their arguments, evaluate evidence, and reflect on their reasoning processes. This comprehensive feedback mechanism supports the development of critical thinking skills, essential for effective argumentative writing among L2 learners.

## 2.3 Exploring and Suggesting a New Model For The Symbiotic Relationship Between ChatGPT, Argumentative Writing, and Critical Thinking Skills

The PECTM model emphasizes key elements of argumentative writing- Position, Explanation, Evidence, and Conclusion-while the Hattie and Timperley model focuses on feedback at task, process, self-regulation, and self levels.
By combining these frameworks, we aim to explore how ChatGPT-generated feedback can support L2 learners in crafting stronger argumentative essays and developing critical thinking skills. Our proposed framework illustrates how these two models align and interact, offering a practical strategy for improving both writing quality and critical thinking through the structured use of ChatGPT. This integrated approach addresses the current research gap and emphasizes ChatCPT's potential as an educational tool.

### 2.3.1 Relationship between ChatGPT, Argumentative Writing, and Critical Thinking: Developing the ChatGPT Guideline for Critical Argumentative Writing (CGCAW) Framework for L2 Learners

Our research has identified the potential for a symbiotic relationship between ChatGPT, argumentative writing, and critical thinking skills. Tan (2023) highlighted that the major components of an argumentative essay inherently involve elements of critical thinking, which are essential for deeper engagement with the material. We used the PECTM model to further support the relationship between argumentative essays and critical thinking, illustrating how components such as analysis, inference, and evaluation are crucial for constructing compelling arguments.

Hattie & Timperley's feedback model (2007) illustrated the significance of feedback in enhancing argumentative writing skills by directing learners' attention to crucial aspects of their writing that need improvement, including clarity in argumentation, structural coherence, and effective evidence use. Our study expands this understanding by demonstrating how ChatGPT can be instrumental in this feedback process, providing real-time, contextual feedback that is critical for learners to revise and improve their essays. ChatGPT's feedback aligns with the four levels of the Hattie & Timperley's model—task, process, self-regulation, and self—helping students understand the alignment of their arguments with the assigned topic, refine their arguments with robust reasoning, and question potential weaknesses.

To effectively address the challenges in crafting high-quality argumentative essays, as supported by Su et al. (2023), the introduction of AI like ChatGPT is essential. By providing structured AI feedback, ChatGPT aids students not only in receiving feedback but also in learning to question their assumptions, critically evaluate their evidence, and refine their arguments. This iterative process fosters deeper critical thinking and enhances argumentative writing skills.

Building on these insights, we propose the ChatGPT Guideline for Critical Argumentative Writing (CGCAW) model (shown in Table 1 below). This new framework integrates the strengths of the PECTM and Hattie & Timperley models to guide L2 learners in utilizing ChatGPT to enhance their argumentative writing while practicing their dialogic critical thinking skills. Our research aims to demonstrate how this structured approach can significantly enhance the capabilities of L2 learners in argumentative writing and foster critical thinking skills.

## 2.3.2 Structured Guidelines for Enhancing Critical Thinking and Argumentative Writing Skills in L2 Learners when using the CGCAW framework

The ChatGPT Guideline Critical Argumentative Writing (CGCAW) will utilize ChatGPT's capabilities as an interactive dialogue partner that not only answers student queries but also challenges their reasoning and argumentation, aiming to foster deeper critical thinking and craft stronger arguments. By using the CGCAW model, we will offer structured guidelines that prompt students to systematically address each component of their argumentative essays

essays—from establishing a clear stance to supporting it with evidence and formulating a persuasive conclusion essay—from establishing a clear stance to supporting it with evidence and formulating a persuasive conclusion.

| Critical Engagement with GPT for language learner | Critical thinking skills | Task Level<br><br>How well tasks are understood/performed (factual) | Process Level<br><br>The main process needed to understand/perform tasks (conceptual) | Self-Regulation Level<br><br>Self-monitoring, directing, and regulating of actions (procedural) | Self Level<br><br>Personal evaluations and affect about the learner (meta-cognitive) |
|---|---|---|---|---|---|
| **Position(s):**<br><br>central position and alternative positions, each with main claim and supporting claims | **Points of View** | What stance could I take for an argumentative essay on the [topic]? | What claims could I develop to construct my argument effectively? | How can I determine if I am confident in the position I've chosen for my argument? | How do I know whether I am confident with the claim I plan on using in my argument ? |
| **Explanation(s):**<br><br>explanations for supporting claims | **Concepts** | X | X | How effectively does the explanation support my argumentative position? | How can I establish a causal connection between my position and its supporting explanation? |
| | **Assumptions** | What is the explanation that supports my position? | Are there any examples/metaphors that support my position? | X | X |
| **Evidence:**<br><br>personal experiences, statistics, anecdotes, research findings, etc. | **Information** | What are the appropriate references, such as statistics, quotations, or news articles, to support my claim? | What structure should I develop to support my claim? | X | X |

| Conclusion: restatement of the position(s), inferences, and implications | Points of View | X | X | How can I reinforce my position to effectively conclude this essay? | X |
|---|---|---|---|---|---|
| | Inferences | X | X | What is the inference process that leads to new understanding? | What new insights can be derived from the argument presented above? |
| | Implications | What would be an appropriate way to conclude this argument? | X | X | X |

Figure 3: ChatGPT Guideline Critical Argumentative Writing (CGCAW) framework

## 3. Methodology

### 3.1 Conducting Experiment [Appendix 1&2]

We decided to experiment with 10-second language learner students from a state university. A random assignment into two groups of five participants ensured an unbiased group composition. Both groups were tasked with writing a short argumentative essay of 150 and 300 words within 40 minutes on the same predetermined topic: *"Is social media beneficial or harmful to society? Discuss the impact of social media on relationships, mental health, and community engagement."*

An evaluation of the CGCAW framework involved providing structured guidelines to the experimental group for their use of ChatGPT. The guidelines provided specific prompts for argument development structure alongside instructions for clarity maintenance and coherence preservation as well as audience awareness and word selection. The control group received no specific instruction when using ChatGPT which enabled researchers to directly assess the quality of their essays.

All participants received detailed instructions through email before the experiment began. The experimental group received comprehensive step-by-step instructions through email (Appendix 1) alongside the control group's basic task description (Appendix 2). Participants received information about task requirements together with time limitations and a requirement to email their finished essays to the experimenter.

The writing environment was monitored through a video conferencing platform during the 40-minute session which enabled compliance verification. Participants needed to stay in the session with their cameras active to follow instructions and work on their assignments independently. The experimental design enabled observation of the writing process without interruption while maintaining participant compliance with experimental requirements.

Each participant sent their finished work to the designated email address for evaluation. We recognize that complete control of writing conditions during the process remained unattainable despite efforts to standardize the environment. Therefore, the validity of the results relies, in part, on the participants' adherence to the instructions, which is discussed in the limitations section of the conclusion.

### 3.2 Argumentative Writing Topic

The participants were given argumentative writing topics specifically chosen for this experiment. Each group employed the distinct strategies outlined earlier: one group will use the CGCAW guidelines while the other group will use ChatGPT without any specific guidelines.

*Q1. Is social media beneficial or harmful to society? Discuss the impact of social media on relationships, mental health, and community engagement.*

### 3.3 Grading Criteria

Once participants finish their argumentative essay writing, they will be evaluated based on the following grading criteria:

- Assessed by the course instructor (40%): A score given based on a clear thesis, evidence and reasoning, addressing counter arguments, and a compelling summary.

- Assessed by Grammarly (30%): A score given based on the correctness, clarity, engagement, and delivery of the essay.

- Assessed by ChatGPT (30%): A score given based on clarity, logical coherence, use of evidence, and persuasiveness.

Initially, the essays written by the participants will be assessed by an instructor with expertise in writing evaluations, contributing 40% of the overall grade. Following this, Grammarly, a cloud-based writing assistant, will evaluate the essays focusing on spelling, grammar, punctuation, clarity, engagement, and delivery, contributing 30% to the grade. Finally, ChatGPT will assist in evaluating and comparing the essays, analyzing factors such as clarity, logical coherence, use of evidence, and persuasiveness, contributing the remaining 30% to the overall assessment.

## 4. Analysis and Outcomes

Before we explore the specifics of the data we collected, it is important to note that Participant 3 and Participant 5 were unable to complete their essays due to unfamiliarity with our CGCAW framework within the given amount of time. Consequently, while their essays were evaluated using the established grading criteria for research purposes, their scores have been excluded from our final calculations. This decision was made to ensure the accuracy and reliability of our data analysis.

### 4.1 Data Collection 1 - Grammarly Assessment

As shown in Figure 4 below, Grammarly assessed the argumentative essays of both Group 1 (experimental) and Group 2 (control). The experimental group (Group 1) had a mean score of 90.6, while the control group (Group 2) achieved a mean score of 92.2. This data indicates a trend where Group 2, which did not use the CGCAW framework, scored higher on the writing assessment than Group 1, which followed the CGCAW framework while using ChatGPT.

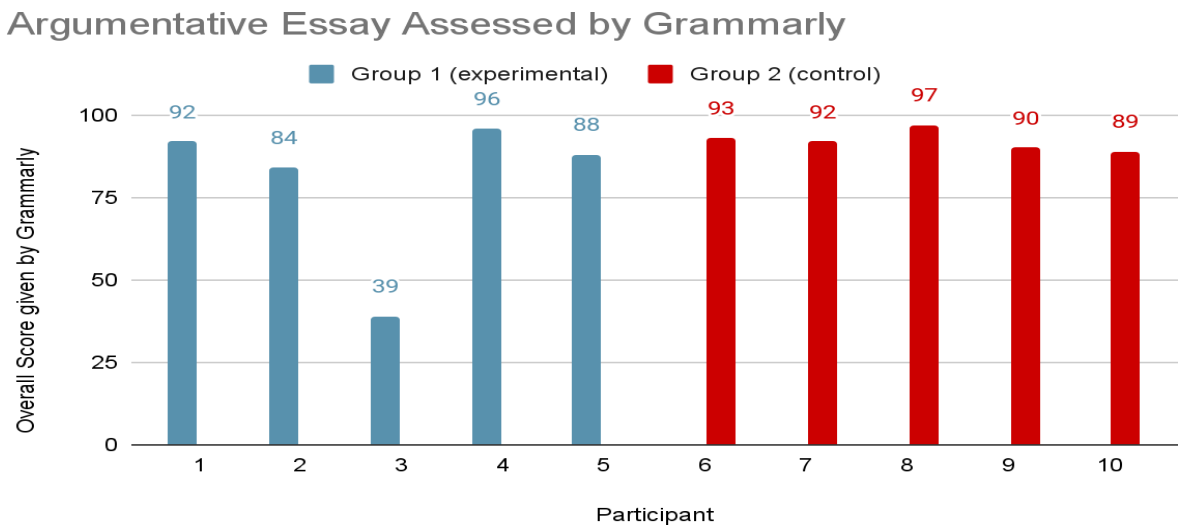

Figure 4: Participants' Argumentative Essay Graded by Grammarly

## 4.2 Data Collection 2 - ChatGPT Assessment

Additionally, we asked ChatGPT to evaluate the argumentative essays of the 10 students, scoring each essay out of 100 based on its clarity, logical coherence, use of evidence, and persuasiveness. ChatGPT was instructed to strictly adhere to these criteria.

Depicted in Figure 5 below, the results showed that the experimental group (Group 1), which used the CGCAW framework, received a mean score of 74%. In contrast, the control group (Group 2), which did not use the framework, received a mean score of 72.2%. The mean score represents an essential metric yet standard deviation analysis helps to determine participant consistency within individual groups. Group 1 participants exhibited moderate score variability with a standard deviation of 7.6 yet Group 2 participants demonstrated slightly better consistency through their standard deviation of 5.5.

The examination of individual assessment areas demonstrates that Group 1 achieved better results than Group 2 across multiple key performance indicators. Statistics reveal Group 1 students averaged 20.3 while showing a standard deviation of 2.4 while Group 2 students obtained an average score of 19.4 with a standard deviation of 1.8. The analysis showed Group 1 participants used evidence at a rate of 17.7 (SD=1.7), which exceeded the rate of 17 (SD=1.2) achieved by Group 2 participants. Participants in Group 1 exhibited logical coherence at 18.4 (SD=1.9) but Group 2 averaged 18 (SD=1.5) in the same category.

The results indicate the CGCAW framework helped students improve clarity, logical coherence, and use of evidence in argumentative essays while showing increased performance variability among Group 1. The measured results show the framework's effectiveness in improving argumentative writing skills yet emphasize the need to refine it for better consistency across user groups.

These results suggest that the CGCAW framework was effective in enhancing the clarity, logical coherence, and use of evidence in argumentative essays. Students who followed the CGCAW guidelines produced higher quality essays in these areas compared to those who did not use the framework. This implies that the CGCAW framework helped students develop better argumentative writing skills in specific areas.

**Argumentative Essay Assessed by ChatGPT**

| Participant | Group | Clarity | Logical Coherence | Use of Evidence | Persuasiveness | Total |
|---|---|---|---|---|---|---|
| Participant 1 | G1 | 21 | 20 | 18 | 19 | 78 |
| Participant 2 | G1 | 20 | 18 | 17 | 17 | 72 |
| Participant 3 | G1 | 15 | 14 | 15 | 14 | 58 |
| Participant 4 | G1 | 20 | 17 | 18 | 17 | 72 |
| Participant 5 | G1 | 16 | 15 | 14 | 15 | 60 |
| Participant 6 | G2 | 22 | 20 | 19 | 20 | 81 |
| Participant 7 | G2 | 20 | 18 | 17 | 18 | 73 |
| Participant 8 | G2 | 19 | 18 | 17 | 17 | 71 |
| Participant 9 | G2 | 18 | 17 | 16 | 17 | 68 |
| Participant 10 | G2 | 18 | 17 | 16 | 17 | 68 |

Figure 5: Participants' Argumentative Essay Graded by ChatGPT

## 4.3 Data Collection 3 - Course Instructor

Lastly, we asked the course instructor from the university to evaluate and assess the argumentative essays. Using the CGCAW framework, the instructor graded the essays based on four criteria: clear thesis, evidence and reasoning, addressing counter arguments, and compelling summary. Each criterion was evaluated on a scale of 100, and the overall score for each essay was calculated.

Indicated in Figure 6 below, Group 1 (experimental group) received a mean overall score of 35, while Group 2 (control group) received a mean overall score of 56. This significant difference suggests that the control group, which did not use the CGCAW framework, performed better in the instructor's assessment.

**Argumentative Essay Assessed by Course Instructor**

| Participant | Group | Clear Thesis (25) | Evidence & Reasoning (25) | Address Counterarguments (25) | Compelling Summary (25) | Total (100) |
|---|---|---|---|---|---|---|
| Participant 1 | G1 | 5 | 5 | 5 | 5 | 20 |
| Participant 2 | G1 | 5 | 15 | 10 | 5 | 35 |
| Participant 3 | G1 | 5 | 5 | 5 | 0 | 15 |
| Participant 4 | G1 | 20 | 10 | 10 | 10 | 50 |
| Participant 5 | G1 | 5 | 5 | 5 | 0 | 15 |
| Participant 6 | G2 | 25 | 25 | 5 | 25 | 80 |
| Participant 7 | G2 | 5 | 10 | 10 | 5 | 30 |
| Participant 8 | G2 | 20 | 5 | 5 | 5 | 35 |
| Participant 9 | G2 | 20 | 20 | 20 | 5 | 65 |
| Participant 10 | G2 | 25 | 25 | 5 | 25 | 70 |

Figure 6: Participants' Argumentative Essay Graded by the Course Instructor

## 5. Discussions

The data collected from Grammarly, ChatGPT, and the course instructor reveal several critical insights into the CGCAW framework's effectiveness in improving argumentative essay writing among second-language learners. Grammarly's assessment showed that the control group (Group 2) slightly outperformed the experimental group (Group 1), suggesting that the CGCAW may need refinement to enhance grammar, punctuation, and clarity. However, ChatGPT's evaluation indicated that the experimental group excelled in clarity, logical coherence, and use of evidence, highlighting the framework's potential to foster critical thinking skills. Contrarily, the course instructor's assessment revealed a significant performance gap favoring the control group, particularly in presenting clear theses and compelling summaries. These mixed results suggest that while the CGCAW improves internal coherence and evidence use, it may initially distract from language mechanics and clear articulation of main arguments. To optimize the CGCAW, focused training, and iterative feedback are recommended to students. Research by Tan (2023) and Noroozi et al. (2016) also suggest the need for long-term studies to evaluate how instructional frameworks influence students' writing skills over an extended period. Building on this idea, it is recommended that future studies examine the CGCAW framework's long-term effectiveness in enhancing argumentative writing among second-language learners. Longitudinal research that incorporates repeated use of the framework and iterative feedback could shed light on its role in fostering sustained writing skill development. Another important consideration is that throughout our experiment, the highest and lowest scores across the three evaluation results indicate limitations in grading writing with AI programs and grammar correction tools. Although AI-based programs have emerged with accurate data, the study should rely more on external instructors who specialize in writing correction rather than solely on course instructors. Additionally, students in the control group outperformed those in the experimental group, suggesting that our CGCAW guideline may not be a suitable template for writing critical argumentative essays with ChatGPT. The fact that students who had the freedom to use ChatGPT without the guideline scored better implies that our guideline needs significant improvement or that a different approach is required to utilize ChatGPT in argumentative writing effectively.

## 6. Conclusion

This study had several limitations that may have affected the accuracy of the results. Firstly, we did not consider the individual capacities of each participant. Despite providing the same instructions to both groups, the different abilities of the participants had a significant impact on the experiment's outcomes. Secondly, participants' prior knowledge of using ChatGPT was not considered. Without familiarity with the tool, some participants might not have utilized it effectively, potentially compromising the experiment's outcomes. Thirdly, the 40-minute time limitations for the essay writing task, proved inadequate for some participants, as indicated by feedback from two participants who needed more time. Lastly, the limited number of participants further constrained the study's accuracy and generalizability. Additionally, due to time constraints, the study was unable to conduct multiple experiments, which would have allowed us to gather more comprehensive data. As a result, we could only conduct a simple experiment, limiting our ability to observe conclusions. This is a significant limitation, as additional experiments would have provided a more reliable basis for evaluating the effectiveness of the CGCAW framework. Although our experiment did not fully meet our expectations, the results reveal significant potential for enhancing the quality of our study. Participants in the experimental group (Group 1) appeared to demonstrate improvements in writing clarity, logical coherence, and use of evidence-skills that are critical for effective argumentative writing. These findings suggest that future research should address the following key areas to more accurately demonstrate the effectiveness of the CGCAW framework.

      Incorporating various assessment tools will provide a more comprehensive evaluation of the CGCAW framework's impact. For example, in addition to Grammarly, ChatGPT, and instructor evaluations, future research could benefit from incorporating assessments by professional writers and peer reviews as additional grading criteria. These additional perspectives would help ensure a wider range of evaluation by highlighting different strengths and weaknesses in the essays, ultimately leading to a more thorough understanding of the framework's effectiveness. Moreover, our study found that Group 1 scored relatively low in persuasiveness, one of the key grading criteria. Future research should investigate how the CGCAW framework can be adjusted to better support the development of persuasive writing skills. The integration of generative AI tools like ChatGPT in second language (L2) learning environments offers promising opportunities for enhancing critical thinking and argumentative essay writing skills. This study explored ChatGPT's efficacy in developing these skills through the structured application of the Paul-Elder Critical Thinking Model (PECTM) and Tan's argumentative writing framework. Our findings indicate that ChatGPT can effectively improve specific aspects of argumentative writing, such as clarity, logical coherence, and the use of evidence. The experimental group, which utilized the ChatGPT Guideline for Critical Argumentative Writing (CGCAW) framework, demonstrated notable improvements compared to the control group. This suggests that the CGCAW framework, when systematically applied, can enhance L2 learners' ability to construct well-argued essays by promoting deeper engagement with critical thinking processes. Notable improvements were observed in the areas of thesis clarity, structured argument development, and the integration of supporting evidence, leading to more coherent and logically consistent essays. However, the study also highlights several limitations and areas for further improvement. The mixed results, particularly the inconsistencies between different assessment methods (Grammarly, ChatGPT, and the course instructor), suggest the need for refinement in the CGCAW framework. These inconsistencies likely arise from the different focus areas of each assessment tool: Grammarly emphasizes grammatical correctness and stylistic elements, while ChatGPT and the instructor may focus more on logical coherence and argumentative strength. Additionally, factors such as individual participant capabilities, prior familiarity with ChatGPT, and the constrained time frame for essay writing should be considered in future research to ensure more accurate and generous results The implications of this research are significant for the field of language education. By using AI tools like ChatGPT, educators can provide more targeted and effective feedback, fostering a more interactive and supportive learning environment. The iterative process of writing and receiving feedback through AI can help students critically evaluate their work and develop stronger argumentative skills.

# Appendix 1

Group1

You have **40 minutes** on this task
Words limit:**150-300 words**

Write about the following topic:

> **Is social media beneficial or harmful to society? Discuss the impact of social media on relationships, mental health, and community engagement.**

Give reasons for your answer and include any relevant examples from your own knowledge and ChatGPT

**Here are guidelines for ChatGPT**
***Must utilize these questions to ChatGPT***

Introduction:

1. What positions can I take to write an argumentative essay for the topic[ ]?
2. What would be the best introduction for an argumentative essay for the topic[ ]? Please be specific

Body paragraph:

1. To what extent does the explanation support my argumentative essay topic [ ]?
2. What would be the right reference including statistics, quotations, news, and so on, that supports the claim for my argumentative essay topic [ ]?
3. What evidence and reasons can support my argumentative essay topic for [ ]?

Conclusion:

1. What would be the right remark to conclude this argument?
2. What would be an appropriate way to summarize my argumentative writing?

**Send to: simon.suh@stonybrook.edu**

# Appendix 2

Group2

You have **40 minutes** on this task
Words limit:**150-300 words**

Write about the following topic:

> **Is social media beneficial or harmful to society? Discuss the impact of social media on relationships, mental health, and community engagement.**

Give reasons for your answer and include any relevant examples from your own knowledge and ChatGPT

You can **freely ask questions to ChatGPT** without any guidelines
You must complete the Essay with this format

Introduction:

Body paragraph:

Conclusion:

**Send to:** simon.suh@stonybrook.edu